\documentstyle[12pt,epsf]{article}
\newcommand{\lesm}
{\ \raisebox{.225ex}{$<$}\hspace*{-1.8ex}\raisebox{-.675ex}{$\sim\ $}}
\newcommand{\gtrm}
{\ \raisebox{.225ex}{$>$}\hspace*{-1.8ex}\raisebox{-.675ex}{$\sim\ $}}

\begin{document}
\title{Energy barrier in the two-Higgs model}

\author{ {\large B. Kleihaus}
\\
{\small Department of Mathematical Physics,
National University of Ireland}\\
{\small Maynooth, Ireland}
}

\date{}

\maketitle
\begin{abstract}
The electroweak model is extended by a second Higgs doublet
and a numerical investigation of static, finite energy classical 
solutions is performed. The results indicate that for a large
domain of the parameters of the Higgs potential, the 
energy barrier between topologically distinct vacua
of the Lagrangian is constituted by a bisphaleron.
\end{abstract}
\medskip \medskip
\newpage
\section{Introduction}
Baryon and lepton number are not strictly conserved 
in the standard model of electroweak interactions \cite{toof}
(see \cite{ru}, \cite{trodden} for recent reviews).
In the attempts to evaluate the rate of baryon number violating
transitions \cite{shap},
 the finite energy barrier between 
inequivalent vacua of the electroweak Lagrangian plays a crucial
role. This energy barrier is characterized by a static classical 
solution of the equations of motion~: the sphaleron \cite{ma}.

The sphaleron was first constructed 
for vanishing Weinberg angle, $\theta_W = 0$,
where the equations of motion
simplify by means of a spherically symmetric ansatz \cite{km}.
Later, the sphaleron solution was constructed for
finite Weinberg angle \cite{kkb}.
There the expectation \cite{km} that the sphaleron energy for
the physical mixing angle $\theta_W \approx 30^{\circ}$ is very close to
the case $\theta_W = 0$ was confirmed. 

In the meantime, it was shown
\cite{bk,kb,yaffe} that the sphaleron constructed by Klinkhamer and Manton
(KM) does not represent the top of the minimal energy barrier 
when the Higgs mass parameter exceeds a critical value.
Indeed, for $M_H > 12 M_W$ the configuration realizing the
minimal energy barrier is given by another static solution
of the equations, the bisphaleron. 
The name refers to the fact, that the bisphaleron solutions appear as a pair,
transformed into each other by parity.
Their classical energy is lower than the one of the KM sphaleron. 

Although interesting from the point of view
of non-linear differential equations admitting bifurcations,
the bisphaleron does not appear to be physically relevant in the (minimal)
one Higgs standard model (1HSM), because
the bisphaleron exists only for a value of the Higgs boson mass 
much higher that the actual upper bound \cite{sophie}
for the Higgs boson mass~: $M_H < 440\ {\rm GeV} \approx 5.5\ M_W$. 

However, several extensions of the minimal electroweak
model are considered as alternative candidates for the
description of the electroweak interactions (see e.g. \cite{hunter}). 
Among these extensions
the ones incorporating a second Higgs doublet in the model (2HSM) play 
a central role, such as the minimal supersymmetric extension,
which is considered for many theoretical reasons.
The upper bound on the Higgs boson mass  $M_H < 440\ {\rm GeV}$ 
is not valid in general in supersymmetric extensions of the Standard Model.
Here upper bounds for the lighter neutral CP-even Higgs bosons are found 
to be in the range $130$ -- $160\ {\rm GeV}$ \cite{sophie}, 
depending on the model under consideration. 
However, these bounds are not valid for the heavier neutral
CP-even Higgs boson.

It is therefore natural to wonder whether the sphaleron-bisphaleron
bifurcation persists in the presence of a second  Higgs doublet
\cite{KPZ} and
to investigate whether the bisphaleron exists
for more physically realistic values
of the parameters. This problem is addressed in detail in this paper
for a family of Higgs potentials depending on four parameters
and leading to a Higgs mechanism for each of the two doublets.
Sphaleron, (winding) bisphaleron and (deformed) bisphaleron 
solutions were first constructed in \cite{btt}.
A detailed numerical analysis of the sphaleron-bisphaleron bifurcation
strongly suggests that, for
a large domain of the parameters of the Higgs potential, the bisphalerons
determine the energy barrier of the model.

In Sect.~2, we present the model, the notations and the
physical parameters.  The spherically symmetric
ansatz, the equations and boundary conditions are given in Sect.~3;
the numerical solutions are then discussed in Sect.~4. 

\section{The model}
We consider the electroweak Lagrangian with two Higgs doublets,
$\Phi_{(1)}, \Phi_{(2)}$, in the limit $\theta_W=0$ (i.~e. $g'=0$).
Using the standard conventions for the covariant derivative
and gauge field strength, the Lagrangian reads
\begin{equation}
{\cal L}
 = -\frac{1}{4} F_{\mu\nu}^a F^{\mu\nu,a}                       
+ (D_\mu \Phi_{(1)})^{\dagger} (D^\mu \Phi_{(1)})                                           
+ (D_\mu \Phi_{(2)})^{\dagger} (D^\mu \Phi_{(2)})
- V(\Phi_{(1)}, \Phi_{(2)}) \ ,                             
\label{lag}
\end{equation}
with
\begin{equation}
\label{cova}
F_{\mu\nu}^a=\partial_\mu V_\nu^a-\partial_\nu V_\mu^a                          
            + g \epsilon^{abc} V_\mu^b V_\nu^c \ ,                                   
\end{equation}
\begin{equation}
D_{\mu} \Phi_{(p)} = \Bigl(\partial_{\mu}                                             
             -\frac{i}{2}g \tau^a V_{\mu}^a                                     
               \Bigr)\Phi_{(p)} \ , p=1,2 \ .                                 
\end{equation}
The most general potential constructed with two Higgs doublets
is presented in \cite{hunter} and depends on nine constants.
Along with Ref.~\cite{btt}
we consider in this paper the family of potentials of the form
\begin{equation}
\label{pot}
     V(\Phi_{(1)}, \Phi_{(2)}) 
    = \lambda_1 (\Phi_{(1)}^{\dagger} \Phi_{(1)} - {v_1^2 \over 2})^2  
    + \lambda_2 (\Phi_{(2)}^{\dagger} \Phi_{(2)} - {v_2^2 \over 2})^2  
\end{equation}
depending on four parameters.
The terms directly coupling the two Higgs doublets are
thus ignored but the main point is that  the above potential 
imposes a Higgs mechanism for each of the Higgs doublets.

The Lagrangian (\ref{lag}) is invariant under SU(2) gauge
transformations and it possesses a large global symmetry 
under    SU(2) $\times$ SU(2) $\times$ SU(2),
which is revealed when using the standard trick
\begin{equation}
     \left(\matrix {\phi_+ \cr \phi_0\cr } \right)
  \longrightarrow  M \equiv
       \left(\matrix {\phi_0^*& \phi_+\cr
                      -\phi_+^* & \phi_0\cr}\right)
\end{equation}
and replacing the Higgs doublets in (\ref{lag}) in terms of the
corresponding matrices $M_1$ and $M_2$.
The new form of the Lagrangian is then manifestly invariant
under the transformations
\begin{equation}
\label{custo}
   V_{\mu}' = A V_{\mu} A^{\dagger} \ \ , \ \ 
   M_1' = A M_1 B  \ \ , \ \ 
   M_2' = A M_2 C\ ,
\end{equation}
 with $A,B,C \in$ SU(2); this is known as custodial symmetry.
Then, the double Higgs mechanism imposed by the potential 
(\ref{pot}) leads to a mass $M_W$ for the three gauge 
vector bosons and to masses $M_{H_1}$, $M_{H_2}$  for
 the surviving degrees of freedom of the  Higgs fields.
In terms of the parameters of the Lagrangian, these masses are
given by
\begin{equation}
\label{masse}
     M_W = {g \over 2} \sqrt{v_1^2 + v_2^2} \ \ \ , \ \ \ 
     M_{H_p} = v_p \sqrt{2 \lambda_p}  \ \ , \ \ p=1,2 \ .
\end{equation}
For later convenience we also define
\begin{equation}
\label{param}
    \tan \beta = {v_2 \over v_1} \ \ , \ \ 
\epsilon_p = 4 {\lambda_p \over g^2} \ \ , \ \ 
\rho_p = {M_{H_p} \over M_W}    \ , \ p=1,2 \ .
\end{equation}
Note, that the classical solutions of the model are characterized by only
three independent parameters, 
which we choose to be
$\beta,$ $\rho_1$, $\rho_2$.
 We consider only $v_1 \geq 0$, $v_2 \geq 0$ so that 
$0 \leq \beta \leq \pi/2$.

\section{Spherical symmetry}
In order to construct classical solutions of the Lagrangian (\ref{lag}),
we employ a spherically symmetric ansatz for the fields.
With the notations of \cite{akiba}, it reads
\begin{eqnarray}
\label{sphsym}
V^a_0 &=& 0\ , \nonumber\\
V^a_i &=& {1-f_A(r)\over{gr}} \epsilon_{aij}\hat r_j + {f_B(r)\over{gr}}
(\delta_{ia}-\hat r_i \hat r_a)+{f_C(r)\over{gr}}\hat r_i \hat r_a
 \ , \nonumber\\
\phi_{(1)} &=& {v_1\over{\sqrt{2}}} 
\left[H(r) + i (\hat r\cdot\vec{\tau})K(r)\right]
\left(\begin{array}{c} 0\\ 1\end{array}\right)
\ , \nonumber\\
\phi_{(2)} &=& {v_1\over{\sqrt{2}}} 
\left[\tilde H(r) + i (\hat r\cdot\vec{\tau})\tilde K(r)\right]
\left(\begin{array}{c} 0\\ 1\end{array}\right) \ .
\end{eqnarray}
The custodial symmetry is used 
to align the two Higgs field parallel to each other
asymptotically.
This spherically symmetric ansatz 
leaves a residual gauge symmetry which can be exploited
to eliminate one of the seven radial functions \cite{bk,yaffe,akiba}. 
Here we adopt the radial gauge
$x_j V^a_j = 0$, which implies $f_C = 0$.

This leads to the following expressions
for the classical  energy  
\begin{eqnarray}
\label{efunc}
   E&=&{M_W \over \alpha_W} \int_0^{\infty} {\cal E} dx \nonumber \\
    &\equiv &{M_W \over \alpha_W} \tilde E \ ,
\end{eqnarray}
with $\alpha_W = {\displaystyle {g^2 \over 4 \pi}}$ and
\begin{eqnarray}
\label{enden}
{\cal E} &=&  \left\lbrace (f'_A)^2+(f'_B)^2+{1\over{2x^2}}
(f^2_A+f^2_B-1)^2\right.\nonumber\\
&+& \cos^2\beta 
\left[ ( H(f_A-1)+Kf_B )^2
      +( K(f_A+1)-Hf_B )^2
      +2x^2\left((H')^2+(K')^2\right)\right. \nonumber\\
& & 
\ \ \ \ \ \ +\left.
     ( \tilde H (f_A-1)+ \tilde K f_B )^2
+    ( \tilde K (f_A+1)-\tilde Hf_B )^2
+ 2x^2\left((\tilde H')^2+(\tilde K')^2\right)\right]\nonumber\\
&+&
\left. 
\cos^4\beta 
\left[\epsilon_1 x^2 \left( H^2+K^2-1\right)^2
     +\epsilon_2 x^2 \left( \tilde H^2+ \tilde K^2-\tan^2\beta\right)^2\right]
     \right\rbrace,
\end{eqnarray}
where the dimensionless variable $x=M_W r$ has been used 
and the prime indicates the derivative
with respect to $x$. 
The equations of motion can then be obtained by varying the functional
(\ref{efunc}) with respect to the six radial functions.
Note, that in the case 
$v_2=0,\lambda_2=0$ the equations for $\tilde H , \tilde K$ decouple 
and these functions can be set consistently to zero,
the remaining system then corresponds to the one of the 1HSM.

Regularity of the solutions at the origin 
imposes the condition $f_A^2 + f_B^2 = 1$ at $x=0$,   
the custodial symmetry (\ref{custo}) can then be exploited to
fix the following values of the radial fields at the origin    
\begin{equation}
\label{condin}
  f_A(0) = 1 \ \ , \ \ f_B(0) = 0 \ \ , \ \ 
  K(0) = 0 \ \ , \ \ \tilde K(0) = 0 \ \ , \ \
  H'(0) = 0 \ \ , \ \ \tilde H'(0) = 0  \ .
\end{equation}
The condition of finiteness of the classical energy imposes
the following asymptotic form on the functions
\begin{eqnarray}
\label{condas}
&(f_A,f_B)_{x=\infty} &= (\cos 2\pi q , \sin 2 \pi q) \ , \nonumber \\
&(H,K)_{x=\infty} &= (\cos \pi(q-k),\sin \pi(q-k))\ ,  \\
&(\tilde H, \tilde K)_{x=\infty} &= 
\tan \beta \ (\cos \pi q,\sin \pi q) \  , \nonumber
\end{eqnarray}
for some real number $q$, and $k$ equal to zero or one.

\section{Numerical results}
In order to make the following discussion self-contained, 
we first summarize the 
main features of the solutions in the 1HSM, i.~e.~the case
$v_2=\lambda_2=0$ with  $\tilde H =  \tilde K = 0$.

\subsection{1HSM}

The KM sphaleron exists for all values of  $\rho_1$ \cite{km}.
For this solution $f_B = H = 0$, and 
the classical energy increases monotonically as a function of $\rho_1$
\begin{equation}
   \tilde E_s(\rho_1=0) \approx 3.04 \ \ , \ \ \tilde E_s(\rho_1=\infty) 
   \approx 5.41 \ .
\end{equation}
The KM sphaleron is always unstable,
but the number of its directions of instability increases
with increasing $\rho_1$, when new solutions, bisphalerons,
bifurcate from the sphaleron \cite{yaffe,bksta}.
These bisphalerons have lower energy than the sphaleron.
At $\rho_1 \approx 12.04$ the first bisphaleron
bifurcates from the sphaleron,
\begin{equation}
   \tilde E_{bs}(\rho_1=12.04) \approx 4.86 \ \ , \ \ 
   \tilde E_{bs}(\rho_1=\infty) \approx 5.07 \ .
\end{equation}

\subsection{2HSM}
Solutions of the Lagrangian (\ref{lag}) were first constructed  in \cite{btt}
(see also \cite{KPZ} for another choice of the Higgs potential).

Confirming the results of \cite{btt}, 
our analysis indicates that a solution of the KM type, i.e.~with
\begin{equation}
          f_B = H = \tilde H = 0
\end{equation} 
exists for all values of the parameters of the potential.
In these solutions the asymptotic angle (\ref{condin})
is always $q=0.5$,
irrespectively of the parameters of the potential.

In Fig.~1 we have plotted the energy of the KM sphaleron for $\rho_1 = 2$
and $\rho_2 = 1$ (i.~e.~when no bisphaleron is present)
as a function of the angle $\beta$. The height of the
barrier does not vary  considerably with $\beta$; changing the
values of $\rho_1,\rho_2$ leads to a very similar picture.

For sufficiently high values of $\rho_1$ 
(or of $\rho_2$) bisphalerons start to exist.

In \cite{btt} it was found that in the special case 
$\rho_1=\rho_2$ deformed bisphalerons bifurcate from the 
sphaleron at the same critical values of the Higgs masses, 
$M_H=12.04 M_W, 128 M_W , ... ,$ as in the 1HSM. 
Indeed, for $\rho_1=\rho_2$ these solutions
can be obtained form the sphaleron and bisphaleron solutions
of the 1HSM by scaling the Higgs field functions $\tilde{K}, \tilde{H}$
by the factor $\cot\beta$ \cite{btt}. 
However, there are additional solutions in the 2HSM bifurcating 
from the sphaleron at values of the Higgs mass as small as 
$M_H \approx 5.5 M_W$
for $\rho_1=\rho_2$ \cite{btt}.
These solutions do not possess counterparts in the 1HSM and, 
like the deformed bisphalerons, they are unstable and 
appear as a pair \cite{btt}.
One characteristic of these bisphaleron solutions in the 
2HSM is that the angle 
$\tilde \phi = \arctan( \tilde K / \tilde H)$ 
increases monotonically from
$0$ (for $x=0$) to $ \pi q$ (for $x= \infty$), while 
$\phi = \arctan( K / H)$ decreases from $0$ to $\pi(q-k)$.
The difference of the angles $\tilde{\phi}-\phi$, 
increases from $0$ to $\pi k$, corresponding to the relative phase
between the two Higgs bosons, which can be characterized by the degree
of a map $S^3\rightarrow S^3$ \cite{btt}. 
These solutions are referred to as winding solutions $W_k$ in \cite{btt}.
In this paper we prefer to use the name winding bisphalerons.
The energy of the winding bisphaleron with $|k|=1$ is lower than the 
energy of the sphaleron or the deformed bisphalerons \cite{btt}. 
Thus, if for given parameters of the model the winding bisphaleron 
with $|k|=1$ exists, it represents the barrier between topologically distinct 
vacua.

In this paper we will restrict ourselves to the lowest energy 
bisphaleron solutions, the winding bisphalerons with $|k|=1$, 
which hereafter also will be called bisphalerons for short, when
no confusion with the deformed bisphalerons can arise. 
For the winding bisphaleron solutions, the six radial functions are non
vanishing and fulfill the boundary conditions (\ref{condas}). 
The asymptotic angle  $q$ depends of the parameters but remains
close to $0.5$
(e.~g.~$q=0.464$ for $\rho_1= 14, \rho_2 = 1, \beta = 0.2$).

We analyzed in detail the sphaleron-bisphaleron transition in the 2HSM. 
For fixed values of $\rho_2$, we determined 
the critical value $\rho_{1,cr}(\beta)$ of the bifurcation
as a function of $\beta$.
The results are summarized in Fig.~2 for $0< \beta < \pi/2$ and
several values for $\rho_2$.

Before we turn to the discussion of the bifurcations in general 
let us discuss first
the case $\rho_1=\rho_2$, which was considered in \cite{btt}.
At the sphaleron-bisphaleron transition the functions 
$H$, $\tilde{H}$ and $f_B$ vanish.
Using the reparameterization $\tilde{K}=\tan \beta \hat{K}$, 
we find in this case
that the angle $\beta$ does no longer 
appear in the differential equations for the functions $K$ and $\hat{K}$.
These equations become identical for $\rho_1=\rho_2$.
Assuming that 
the solutions of these equations are unique (up to sign), 
we find the relation $\hat{K}=-K$, 
which is consistent with the boundary conditions 
(\ref{condas}). 
With this relation we find 
that in the differential equation for the function $f_A$
the terms involving the angle $\beta$ cancel.
Thus, 
the angle $\beta$ disappears completely from the differential equations. 
Consequently, 
the critical value $\rho_{1,cr}(\beta)$ is a constant in this case, 
$\rho_{1,cr}(\beta)=\rho_{2,cr} \approx 5.585$. 
The constant $\rho_{2,cr}$ will play a crucial role in the following
discussion of the bifurcations in the general case.

In Fig.~2 we exhibit the critical value $\rho_{1,cr}(\beta)$ 
as a function of $\beta$ for fixed values of $\rho_2$.
The solid curves correspond to $\rho_2< \rho_{2,cr}$ and
the dashed curves correspond to $\rho_2> \rho_{2,cr}$.
For $\rho_2=\rho_{2,cr}$ the critical value $\rho_{1,cr}(\beta)$ 
consists of two intersecting curves (dashed-dotted).
While the first one is a constant given by 
$\rho_{1,cr}(\beta)=\rho_{2,cr}$, 
the second one 
is a monotonically decreasing function of $\beta$, 
which possesses the value $\rho_{1,cr}=12.04$ at $\beta=0$ 
and vanishes at $\beta=\beta^{cr} \approx 0.53$. 
The curves intersect at at $\beta=\beta^*\approx 0.32$.
From Fig.~2 we see,
that the curves with $\rho_2<\rho_{2,cr}$ cannot 
be deformed continuously into the curves with $\rho_2>\rho_{2,cr}$. 

Turning to the question for which ranges of parameters bisphalerons exist, 
we first consider $\rho_2< \rho_{2,cr}$. 
In this case bisphalerons exist for parameters 
$\rho_1 < \rho_{1,cr}(\beta)$ (lower solid curves)
and for 
$\rho_1 > \rho_{1,cr}(\beta)$ (upper solid curves).
For $\rho_1$ below the lower solid curves, 
bisphalerons only exist for restricted ranges of the parameters 
$\rho_2$ and $\beta$, 
i.~e. $3.2 \lesm \rho_2 <\rho_{2,cr}$, $ \beta \leq  \beta_0(\rho_2)$,
where $\beta_0(\rho_2)$ denotes the value of $\beta$ 
for which the critical value $\rho_{1,cr}$ becomes zero 
for a fixed value of $\rho_2$.
We find $\beta_0(\rho_2\approx 3.2)=0$ and 
$\beta_0(\rho_2=\rho_{2,cr})=\beta^{cr} \approx 0.53$.
For $\rho_1$ above the upper solid curves, in contrast, 
the solutions exist for all $\beta$ 
and even for vanishing $\rho_2$.
In general we observe from Fig.~2, 
that with increasing $\rho_2$ 
the bisphalerons exist in an increasing region of the parameter space.
Note, that for $\rho_2< \rho_{2,cr}$ 
the critical value $\rho_{1,cr}(\beta)$ 
is a monotonically decreasing function of $\beta$. 
This is in contrast to the case $\rho_{2,cr}<\rho_2$, 
which we will discuss next.

For $\rho_{2,cr} < \rho_2 \lesm 12.04$ 
the bisphalerons exist outside the regions enclosed by the 
upper and lower dashed curves in Fig.~2.,
i.~e. they exist below and to the right of the upper dashed curves, 
and above and to the left of the lower dashed curves. 
The regions enclosed by the upper dashed curves 
shrink in size with increasing $\rho_2$,  
degenerate to a single point 
($\beta=0$, $\rho_1\approx 12.04$) at $\rho_2\approx 6.0$
and cease to exist for $\rho_2 \gtrm 6.0$.
Similarly, the regions enclosed by the lower dashed curves 
shrink in size with increasing $\rho_2$.
As $\rho_2$ approaches the value $12.04$ 
these regions degenerate to the vertical line 
($\beta=\pi/2$, $0 < \rho_{1,cr}\lesm 6.0$). 
For $\rho_2 \gtrm 12.04$ 
the bisphalerons exist for all values of $\beta$ and $\rho_1$.
Again we find, 
that for increasing $\rho_2$ 
bisphalerons exist in an increasing region of the parameter space.

From the discussion above, 
we find for the case $\rho_2=\rho_{2,cr}$
that bisphalerons exist for 
$\rho_1< {\rm min}[\rho_{2,cr}, \rho_{1,cr}(\beta)]$
and for 
$\rho_1> {\rm max}[\rho_{2,cr}, \rho_{1,cr}(\beta)]$,
where
$\rho_{1,cr}(\beta)$ is given by the dashed-dotted curves in Fig.~2.

To present this complicated situation in a more transparent way,
we exhibit in Fig.~3 a three dimensional plot of the parameter space, 
where the bisphalerons exist outside the shaded region. 
The surface of this region is formed by the set of bifurcation points. 
The curves of Fig.~2 
correspond to the intersections of horizontal slices
with the surface of the shaded region.
Note, that there is a discrete symmetry in Fig.~3, 
given by the transformation 
$\beta \rightarrow \pi/2-\beta$, 
$\rho_1 \leftrightarrow \rho_2$. 
Indeed, this symmetry is also a symmetry of the differential equations.
For any solution $(f_A, f_B, H, K, \tilde{H}, \tilde{K})$
for the parameters 
$(\beta, \rho_1,\rho_2)$ 
there is a solution for parameters 
$(\bar{\beta}=\pi/2-\beta, \bar{\rho}_1=\rho_2,\bar{\rho}_2=\rho_1)$ 
given by 
$(\bar{f}_A= f_A, \bar{f}_B= f_B, 
\bar{H}=-\tilde{H}/\tan \beta, 
\bar{K}=-\tilde{K}/\tan \beta, 
\bar{\tilde{H}}=-H/\tan \beta, 
\bar{\tilde{K}}=-K/\tan \beta)$.
Both solutions possess the same energy 
and obey the same boundary conditions Eq.~(\ref{condas}) 
with the same values of the constants $q$ and $k$. 

The bifurcation of the winding bisphaleron in the 2HSM 
differs considerably from the bifurcations of the bisphalerons 
in the 1HSM, 
where the solutions depend only on one parameter 
and only one bifurcation point exist 
for the lowest energy bisphaleron.
In the 2HSM the number of bifurcation points 
for fixed values of $\rho_2$ and $\beta$
can be zero, one or two.

For fixed $\rho_2\lesm 3.2$ and $0<\beta<\pi/2$ 
there is only one bifurcation point.
 
A second bifurcation point appears if $3.2\lesm\rho_2\lesm 6.5$.
In this case we observe from Fig.~2, 
that for $\rho_2< \rho_{2,cr}$ 
and small values of $\beta$ 
the two bifurcation points 
lie on the upper and lower solid curve, respectively.
With increasing $\beta$ both bifurcation points decrease. 
At the critical value $\beta_0(\rho_2)$ 
the bifurcation point on the lower solid curve vanishes 
and only one bifurcation point remains.
For fixed $\rho_2> \rho_{2,cr}$ 
the bifurcation points lie on the upper and lower dashed curves in Fig.~2.
In contrast to the case $\rho_2< \rho_{2,cr}$ however, 
both bifurcation points 
lie either on the upper dashed curve 
or on the lower dashed curve.
If $\beta$ is small, 
the two bifurcation points belong both to the upper dashed curve.
With increasing $\beta$ the lower bifurcation point increases, 
whereas the upper bifurcation point decreases.
At a critical value of $\beta$ 
the bifurcation points merge into a single point 
and do not exist for larger values of $\beta$. 
As $\beta$ reaches a second critical value, 
another bifurcation point appears, 
which now belongs to the lower dashed curve.
This bifurcation point splits into a pair of points as $\beta$ becomes larger.
The upper bifurcation point of this pair increases with increasing $\beta$
and exist up to $\beta=\pi/2$. 
The lower bifurcation point decreases with increasing $\beta$ 
and vanishes at $\beta=\beta_0(\rho_2)$.

For $6.0 \lesm \rho_2 \lesm 12.04$ there is no bifurcation point for
$\beta<\beta_0(\rho_2)$ and only one bifurcation point for
$\beta>\beta_0(\rho_2)$, 
which lies on the lower dashed curves in Fig.~2.

For $\rho_2>12.04$ no bifurcations of the winding bisphalerons
with winding number $|k|=1$ were found for $0<\beta<\pi/2$.

We demonstrate these bifurcations for fixed 
$\rho_2=5.6$ and for $\beta=0.2$ and $\beta=0.4$
in Fig.~4, 
where the energies of the sphaleron and the bisphaleron
are shown as functions of the parameter $\rho_1$.
For small values of $\rho_1$ 
the bisphaleron (solid and dashed lines) 
and the sphaleron (dotted lines) both exist. 
The energy of the bisphaleron is smaller than the energy of the sphaleron. 
However, the difference of the energies is very small.
When $\rho_1$ reaches the critical value $\rho_{1,cr}(\beta)$, 
the energy curve of the bisphaleron 
merges into the energy curve of the sphaleron. 
For larger values of $\rho_1$, 
bisphalerons do not exist 
until $\rho_1$ reaches the second critical value.
At this point 
the energy curve of the bisphaleron 
bifurcates from the energy curve of the sphaleron. 
For larger values of $\rho_1$ 
both sphaleron and bisphaleron exist 
and the energy of the bisphaleron 
is again less than the energy of the sphaleron.

The curves for $\beta=0.2$ and $\beta=0.4$ look very similar.
However, 
there is a complicated transition as $\beta$ varies 
between the values $\beta=0.2$ and $\beta=0.4$, 
not shown in Fig.~4.
As $\beta$ becomes gradually larger 
the bifurcation points on the energy curve approach each other 
and merge into one single point 
at some critical value of $\beta$. 
At this value of $\beta$ 
the bifurcation point is a tangential point of the two energy curves.
With increasing $\beta$ 
the energy curves separate first 
and approach each other again,
when $\beta$ approaches a second critical value.
At this value of $\beta$
a tangential point is formed again,
which splits into a pair of bifurcation points with increasing $\beta$.

In addition to the sphaleron-bisphaleron transitions 
the 2HSM also possesses transitions of sphalerons and bisphalerons 
to the sphalerons and bisphalerons of the 1HSM. 
These transitions correspond 
to the limits $\beta \rightarrow 0$ (and $\beta \rightarrow \pi/2$).
We will consider two different cases for the limit $\beta \rightarrow 0$. 
In the first case we consider  $\beta \rightarrow 0$ with fixed mass $M_{H_2}$.
In the second case we consider $\beta \rightarrow 0$ 
with $M_{H_2}\rightarrow \infty$ but with fixed $\lambda_2 \tan^4\beta$, 
such that the contribution to the energy of the ($\tilde{K},\tilde{H}$) 
dependent part of the Higgs potential is (at most) finite. 
In both cases we assume that the functions  $\tilde{K}(r)$ and $\tilde{H}(r)$ 
vanish $\sim\tan\beta$ and reparameterize 
$\tilde{K}(r)$ and $\tilde{H}(r)$ as 
$\tilde{K}(r)=\tan\beta \hat{K}$, $\tilde{H}(r)=\tan\beta \hat{H}$,
respectively. The purpose of this parameterization is that we can take the 
limit $\beta\rightarrow 0$ in a well defined way.

For the discussion of the limit $\beta \rightarrow 0$ in the first case, 
we substitute 
$\epsilon_2 = \frac{M^2_{H_2}}{2M_W^2\sin^2\beta}= \frac{\rho_2}{2\sin^2\beta}$
into the differential equations and keep $\rho_2$ fixed.
Then the differential equations for the functions $\hat{K},\hat{H}$
do not contain any explicit $\beta$ dependence.
In addition the boundary conditions at infinity 
for the functions $\hat{K},\hat{H}$ are independent of $\beta$.
In the limit $\beta \rightarrow 0$
the differential equations for the functions $f_A, f_B,K,H$ 
are independent of $\hat{K},\hat{H}$ 
and reduce to the equations of the 1HSM 
with Higgs mass parameter $\rho_1$. 
In the equations for the functions $\hat{K},\hat{H}$ 
the gauge field functions  $f_A, f_B$ are still present.
Thus, the latter system of differential equations 
has to be solved for fixed functions $f_A, f_B$,
The solutions of this differential equations depend on the parameter $\rho_2$ 
and indirectly also on $\rho_1$. 

Let us now discuss the limit $\beta \rightarrow 0$
for the bisphaleron solutions of the 2HSM, 
i.~e.~for parameters  $\rho_1$ and $\rho_2$
outside the shaded region in Fig.~3 at $\beta=0$.
In the case $\rho_1 > 12.04$ 
the functions $f_A, f_B,K,H$ 
approach the corresponding functions of the bisphaleron of the 1HSM,
as $\beta \rightarrow 0$.
In the case $\rho_1 < 12.04$ however, 
the bisphaleron of the 1HSM does not exist 
and the bisphalerons of the 2HSM 
reach the sphaleron of the 1HSM.
Consequently, 
the functions $f_B$ and $H$ vanish in this limit.
For the decoupled equations for $\hat{K},\hat{H}$ 
we found in both cases, 
that they possess nontrivial solutions for all values of $\rho_2$
considered as above. 
In particular, for $\rho_1 < 12.04$ 
the function $\hat{H}$ is nontrivial,
as long as the value of $\rho_2$ 
lies above the shaded region in Fig.~3 at $\beta=0$,
and $\hat{H}$ vanishes as $\rho_2$ approaches the shaded region.
In the limit $\beta \rightarrow 0$ 
the functions $\hat{K},\hat{H}$ do not contribute to the energy. 
Thus, in this limit 
the energy of the bisphalerons of the 2HSM 
is given by the energy of the bisphalerons of the 1HSM 
with Higgs mass parameter $\rho_1$
if $\rho_1> 12.04$, 
and 
by the energy of the sphaleron of the 1HSM if $\rho_1< 12.04$.

For the discussion of the limit $\beta \rightarrow 0$ in the second case,
we again substitute 
$\epsilon_2 = \frac{M^2_{H_2}}{2M_W^2\sin^2\beta}= \frac{\rho_2}{2\sin^2\beta}$.
In this case we fix $\frac{\rho_2}{\sin^2\beta}=const.$ As a
consequence the $\beta$ dependence does not disappear from the differential
equations. 
However, this case is analog to the case discussed before
in the limit $\rho_2 \rightarrow \infty$.
Taking both limits, $\beta\rightarrow 0$ and $\rho_2 \rightarrow \infty$,
simultaneously,
we find again that the energy and the functions 
$f_A, f_B, K,H$ of the 2HSM approach the corresponding energy and functions 
of the bisphaleron of the 1HSM for $\rho_1> 12.04$, 
whereas for $\rho_1< 12.04$ they approach the corresponding 
energy and functions of the sphaleron of the 1HMS. 
In this limit the functions $\hat{K}$ and $\hat{H}$ remain nontrivial 
for $\rho_1> 12.04$ as well as for $\rho_1< 12.04$.
Note, that for $\rho_2 \rightarrow \infty$ the potential for the Higgs 
field functions 
$\hat{H}, \hat{K}$ becomes a constraint, $\hat{H}^2+\hat{K}^2=1$,
which can be fulfilled if we parameterize $\hat{H}=\cos\alpha(r)$, 
$\hat{K}=\sin\alpha(r)$, where $\alpha(r)$ is a function to be determined
by the residual differential equation, which still depends on the 
gauge field functions $f_A$ and $f_B$. 
Once the constraint $\hat{H}^2+\hat{K}^2=1$ is fulfilled,
the contribution of $\hat{H}, \hat{K}$ to the potential vanishes,
which leads to a vanishing contribution to the energy  
in the limit $\beta\rightarrow 0$.

For the discussion of the limit $\beta \rightarrow \pi/2$ 
we can exploit the discrete symmetry stated above. 
If we replace $\beta$  by $\pi/2-\beta$, 
the result is in complete analogy 
to the discussion of the limit $\beta \rightarrow 0$, 
only the roles of $K,H,\rho_1$ and $\hat{K},\hat{H},\rho_2$ are interchanged.

Next we address the question as to what are the lowest possible masses
$M_{H_1}= \rho_1 M_W$, $M_{H_2}= \rho_2 M_W$
of the two Higgs bosons, for which bisphaleron solutions exist. 
In the following we will fix $M_{H_2}$ and determine the lowest
possible value of $M_{H_1}$. 
A discussion of the lowest possible masses of the 
lighter and heavier Higgs bosons is given below.

First we consider the case with fixed $M_{H_2}$ and assume $M_{H_2}<M_{H_1}$.
For small values of $M_{H_2}$, $0< M_{H_2} \lesm 3.2 M_W$, 
the lowest possible mass  $M^{(min)}_{H_1}$
can be determined by the upper solid curves in Fig.~2. 
At $\beta=0$ the minimal value is $M^{(min)}_{H_1}\approx 12.04 M_W$. 
It decreases with increasing $\beta$ 
and becomes as small as $M^{(min)}_{H_1}\approx 3.2 M_W$ 
for $M_{H_2}=0$ as $\beta$ approaches the value $\pi/2$. 
When $M_{H_2}$ becomes larger than $\approx 3.2 M_W$, 
new solutions emerge for small values of $\beta$, 
indicated by the lower solid curves in Fig.~2. 
However, for these bisphalerons  $M_{H_1}$ 
is always smaller then $M_{H_2}$. 
If $M_{H_2}$ becomes larger then the critical mass $\approx 5.585 M_W$, 
the lowest possible mass $M^{(min)}_{H_1}$ 
is again always smaller then $M_{H_2}$. 

We now assume that $M_{H_2}>M_{H_1}$.
As long as $M_{H_2}\lesm 3.2 M_W$ we find $M_{H_1}>M_{H_2}$. 
This case was discussed in the previous paragraph.
For $ 3.2 M_W$ $\lesm$ $M_{H_2}$ $\lesm$  $5.585 M_W$ 
bisphalerons with a vanishing Higgs mass $M_{H_1}$ exist
for a finite range of the parameter 
$\beta$, $0 \leq \beta \leq\beta_0(\rho_2)$ 
(below the lower curves in Fig.~2).
If $\beta$ becomes larger then $\beta_0(\rho_2)$,
only bisphalerons for Higgs masses $M_{H_1}>M_{H_2}$ exist 
(see previous paragraph).
In the case  $M_{H_2}\gtrm 5.585 M_W$ 
bisphalerons for $M_{H_1}=0$ again exist 
for parameters $\beta< \beta_0(\rho_2)$. 
In contrast to the case $M_{H_2}\lesm 5.585 M_W$ however, 
bisphalerons with $M^{(min)}_{H_1}<M_{H_2}$
now also exist for $\beta> \beta_0(\rho_2)$.
In this case the lowest possible mass $M^{(min)}_{H_1}$
for the bisphaleron solutions
is given by the lower dashed curves in Fig.~2.

From the lower dashed curves 
we determine the behavior of 
$M^{(min)}_{H_1}$ as a function of $\beta$,
assuming $M_{H_2}>M_{H_1}$.
For $3.2 M_W\lesm M_{H_2}\lesm 5.585 M_W$ we find 
$M^{(min)}_{H_1}(\beta)=0$ 
for  $\beta < \beta_0(\rho_2)$,
and that no bisphaleron solutions with $M_{H_1}<M_{H_2}$ exist 
for $\beta > \beta_0(\rho_2)$. 
For $5.585 M_W \lesm M_{H_2} \lesm 6.5 M_W$ 
bisphaleron solutions with $M_{H_1}<M_{H_2}$ exist for 
all values of $\beta$. 
In this case the function 
$M^{(min)}_{H_1}(\beta)$ vanishes for $\beta<\beta_0(\rho_2)$ only. 
At $\beta=\beta_0(\rho_2)$ the function $M^{(min)}_{H_1}(\beta)$ 
develops a discontinuity 
as it jumps to the non vanishing value of the lower dashed curve.
For $\beta>\beta_0(\rho_2)$, 
$M^{(min)}_{H_1}(\beta)$ is a monotonically increasing function.
If $M_{H_2}$ is large enough, 
the lower dashed curves in Fig.~2 
are monotonically increasing functions of $\beta$. 
Consequently, 
the discontinuity in the function $M^{(min)}_{H_1}(\beta)$ disappears.
If the Higgs mass $M_{H_2}$ becomes larger than $\approx 12.04 M_W$, 
bisphalerons exist for arbitrary values of $M_{H_1}$ and $\beta$.

From a phenomenological point of view it might be interesting
to discuss the region of existence of the bisphalerons in terms
of the masses of the lighter and heavier Higgs boson, 
$M_h={\rm min}\{M_{H_1},M_{H_2}\}$ and $M_H={\rm max}\{M_{H_1},M_{H_2}\}$,
respectively.
We now will address the question as to what is the lowest possible
value of $M_H$ for fixed $M_h$ and $\beta$, such that bisphaleron 
solutions exist.
First we observe from Figs.~2 and 3, 
that for any $M_{H_2}> \rho_{2,cr} M_W$ we can find some $M_{H_1}<M_{H_2}$,
such that a bisphaleron solution exist, and vice versa, with 
$M_{H_1}$ and  $M_{H_2}$ interchanged.
Thus, for the discussion we have to take $M_h\leq \rho_{2,cr} M_W$.
We further observe form Fig.~2 that for the upper solid lines 
$\rho_2<\rho_{2,cr}$ and $\rho_{1,cr}(\rho_2,\beta)\leq \rho_2$ holds,
i.~e. $M_h=\rho_2 M_W$, $M_H \geq\rho_{1,cr} M_W$.
Now, due to the symmetry $\beta \rightarrow \pi/2-\beta$, 
$M_{H_1}\leftrightarrow M_{H_2}$, it is sufficient to consider 
$M_h = M_{H_2}$ for $\beta\geq \pi/4$
(which is equivalent to $M_h = M_{H_1}$ for $\beta\leq \pi/4$). 
Then the lowest possible mass of the heavier Higgs boson 
$M^{(min)}_H(M_h,\beta)$ can be determined from the upper solid curves in 
Fig.~2 for $\beta\geq \pi/4$. 
For  $\beta\leq \pi/4$ the corresponding mass is given by 
$M^{(min)}_H(M_h,\beta)=M^{(min)}_H(M_h,\pi/2-\beta)$.
For example, fixing $M_h = M_{H_2}=M_W$ we find from Fig.~2
$M^{(min)}_H(M_W,\beta=0)=M^{(min)}_H(M_W,\beta=\pi/2)\approx 4.4 M_W$, 
$M^{(min)}_H(M_W,\beta=\pi/4)\approx 7.7 M_W$.

Let us address the question of the lowest possible masses of 
the two Higgs bosons for the bisphalerons in the limit when the 
2HSM reduces to the 1HSM, in which an upper bound for the 
mass of the Higgs boson exist, 
$M_h< 440 {\rm GeV} \approx 5.5 M_{\rm W}$ \cite{sophie}.
We assume that in this limit the same upper bound also applies to the lighter
Higgs boson in the 2HSM.
For the bisphalerons
in the limit $\beta \rightarrow 0$ 
the mass of lighter Higgs boson 
is given by $M_h=\rho_1 M_W$, if $M_h<440 {\rm GeV}$. 
From Fig.~3 we see that for the bisphalerons the 
lowest possible mass of the heavier Higgs boson $M_H^{(min)}$ 
increases  with increasing $M_h$.
For an (unrealistic) massless lighter Higgs boson we find 
$M_H^{(min)}= 260 {\rm GeV}$, while for a more realistic value
$M_h=90 {\rm GeV}$ \cite{lowbound} 
we find $M_H^{(min)}= 350 {\rm GeV}$.
For the upper bound $M_h=440 {\rm GeV}$ the lowest possible mass
of the heavier Higgs boson becomes $M_H^{(min)}= 450 {\rm GeV}$.
In this case bisphaleron solutions exist, for which the both Higgs bosons
possess roughly the same mass.
In the limit $\beta \rightarrow \pi/2$ the discussion will lead to the 
same result.

\section{Conclusions}

We have analyzed in detail the bifurcation of the first 
winding bisphaleron solution
from the sphaleron solution in the 2HSM.
Indeed, we find that the bisphaleron dominates
the energy barrier between non-equivalent vacua in a non-negligible
domain of the parameter space of the Higgs potential (\ref{pot}).
We discussed the lowest possible masses of the both Higgs bosons,
for which the winding bisphalerons exist.

It would be interesting to extend this analysis to more general
potentials for the two Higgs doublets.
When all the terms of the potential  are positive, it is clear that
the particular case treated here imposes the same
conditions on the Higgs degrees of freedom, namely the Higgs
mechanism for each of the two Higgs doublets.  
As a consequence
all the other degrees of freedom of the Higgs field (especially the ones
describing classically the charged Higgs bosons \cite{hunter})  will be 
forced to vanish asymptotically.
Sphalerons in models with more general potentials have been constructed 
before, taking into account one loop corrections and finite temperature
effects for the potential as well as CP-violating terms,
see e.~g. \cite{KPZ,MOQ,GrantHind}. It was shown
that the energy of the sphaleron does not depend strongly on the
parameters of the potential.
We guess that a similar result will hold for
the energy of the bisphaleron solutions in these models. 
However, the sphaleron-bisphaleron transition may depend strongly on 
the potentials.
The possibility of the dominance of the bisphalerons 
at physically relevant values of the potential parameters should be
considered.

\bigskip

{\bf Acknowledgments}
The author thanks Y. Brihaye and J. Kunz for 
helpful discussions.
This work was carried out under 
Basic Science Research project SC/97/636 of
FORBAIRT.

\newpage
\small{

 }

\newpage
\normalsize

\centerline{Figure Captions}
\begin{itemize}
\item [Figure 1]
The sphaleron energy as a function of the mixing angle $\beta$
for $\rho_1 = 2$, $\rho_2 = 1$. 
\item [Figure 2]
The value $\rho_{1,cr}$ of the sphaleron-bisphaleron bifurcation
of the 2HSM as a function of $\beta$ for several values of $\rho_2$. 
The numbers at the curves denote the value of $\rho_2$.
Bisphaleron solutions exist above the upper solid curves, below 
the lower solid curves and outside the regions enclosed by the 
dashed curves. The star $\ast$ indicates a saddle point.
\item [Figure 3]
For the  2HSM the subset in parameter space, where no bisphaleron 
solutions exist, is shown as the shaded region. 
Its surface is formed by the points
where the sphaleron-bisphaleron bifurcation occur.
The star $\ast$ indicates a saddle point.
\item [Figure 4]
The energy of the sphaleron and bisphaleron of the 2HSM is shown
as a function of the parameter $\rho_1$ for fixed parameters $\rho_2=5.6$ and
$\beta=0.2$ and $\beta=0.4$. The solid (dashed) line correspond to the
bisphaleron solutions  for $\beta=0.4$ ($\beta=0.2$). The dotted lines
correspond to the sphaleron solutions.
The lower and upper bifurcation points for $\beta=0.4$
are indicated by the symbols 
{\bf\footnotesize )} and {\bf\footnotesize (}, respectively, 
 and for $\beta=0.2$ by the symbols 
{\bf\footnotesize ]} and {\bf\footnotesize [}, respectively.
  
\end{itemize}

\newpage

\begin{figure}
\centering
\mbox{\epsfysize=12.cm\epsffile{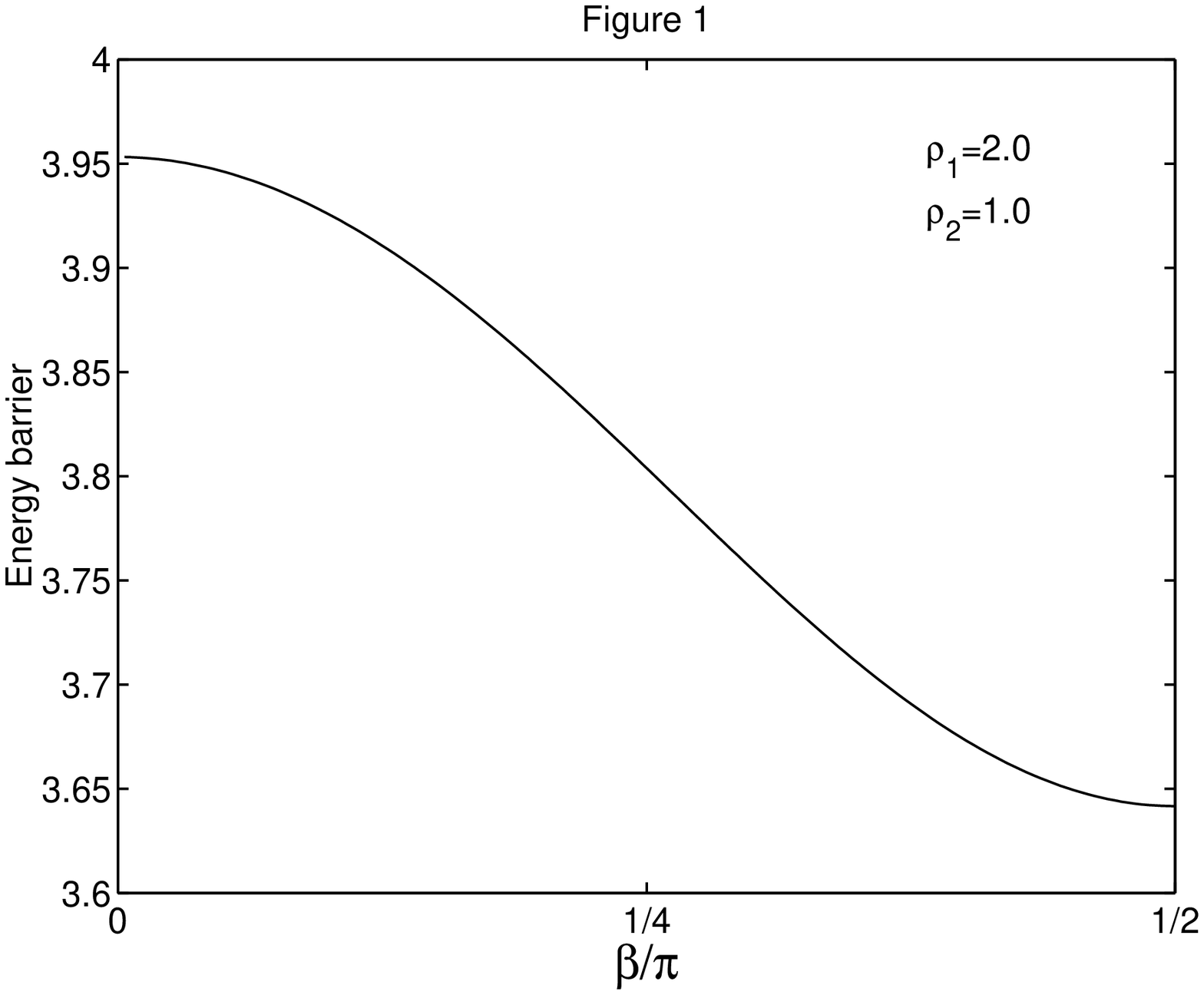}}
\end{figure}

\newpage
\begin{figure}
\centering
\mbox{\epsfysize=12.cm\epsffile{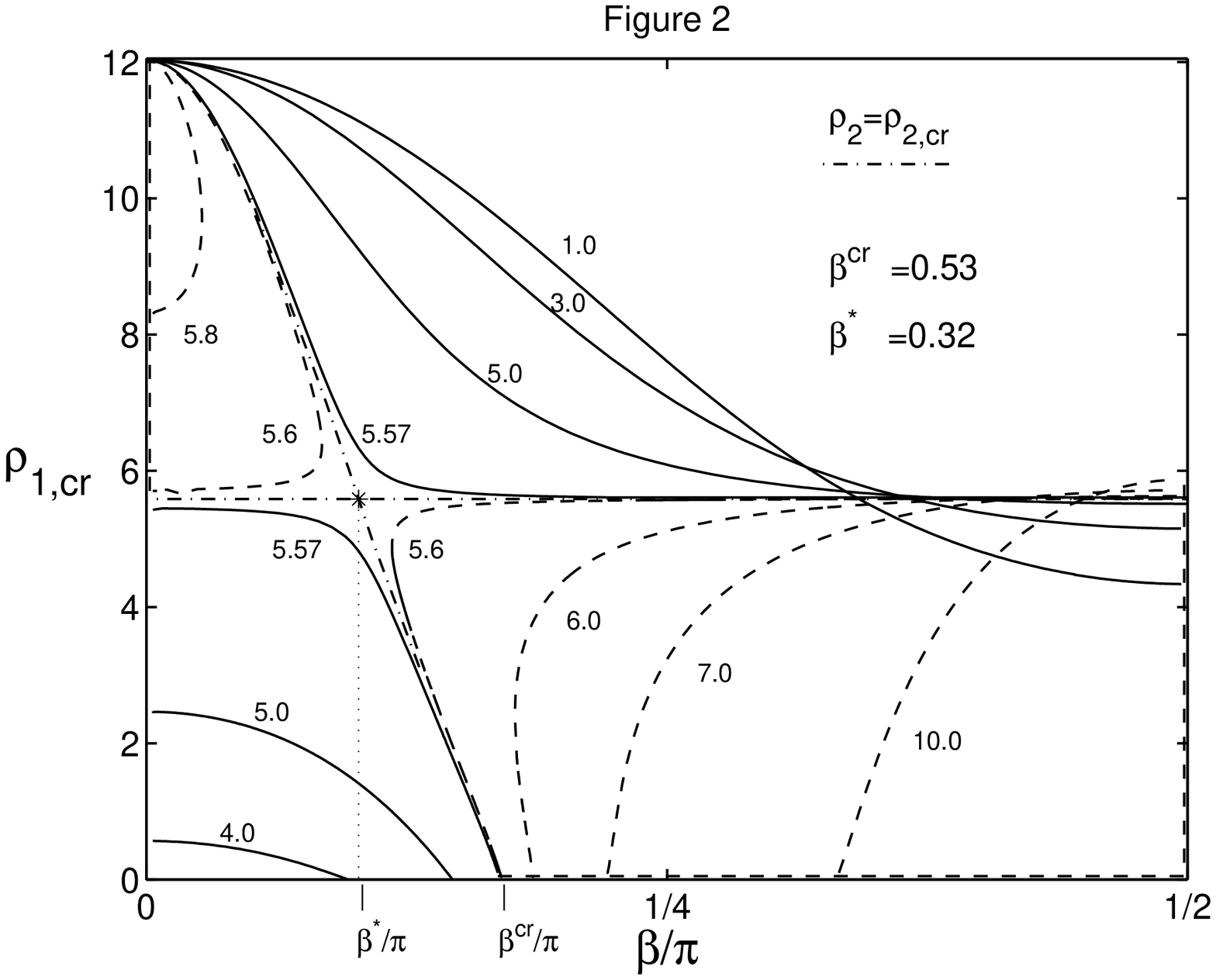}}
\end{figure}

\newpage
\begin{figure}
\centering
\mbox{\epsfysize=12.cm\epsffile{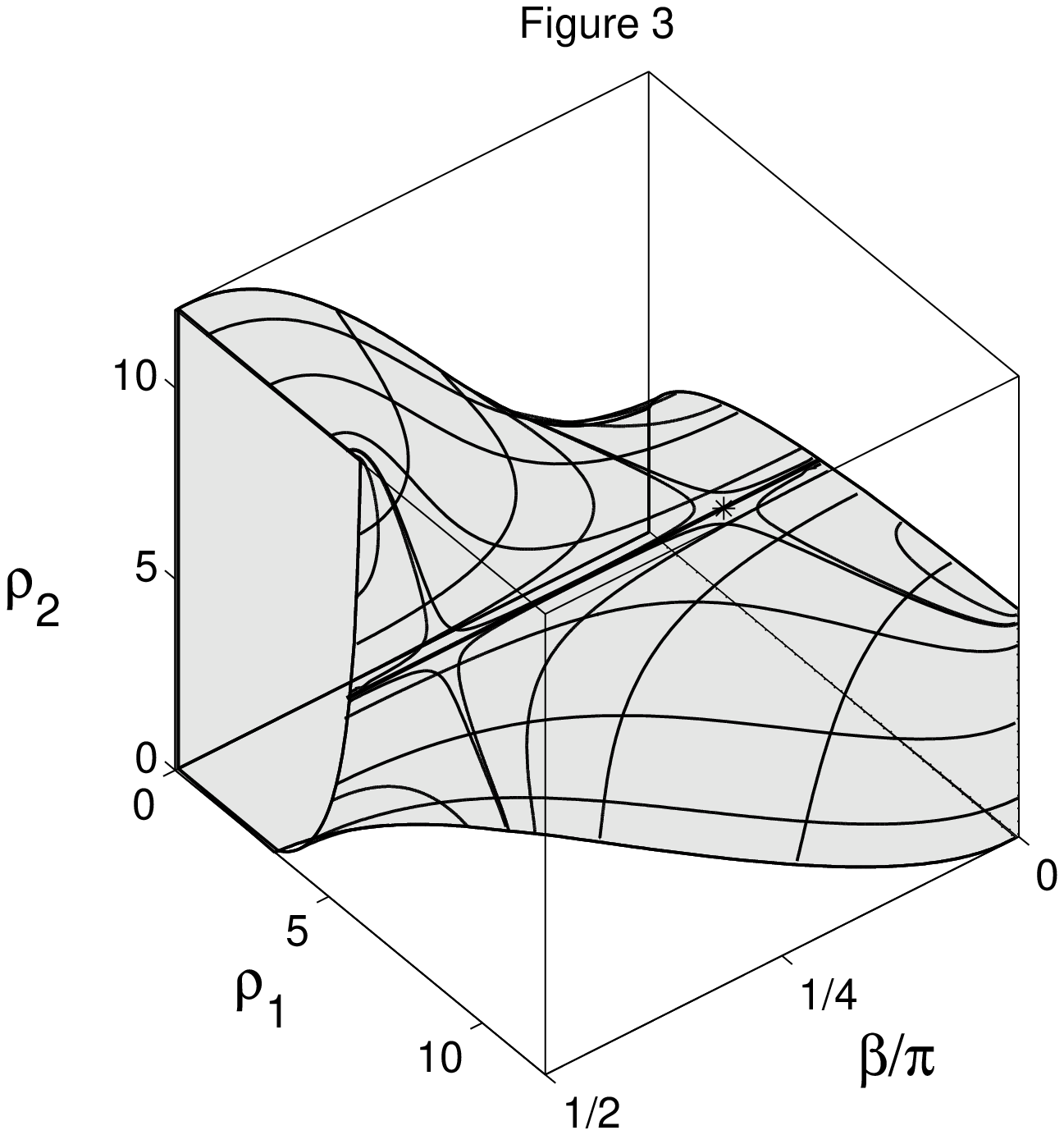}}
\end{figure}

\newpage
\begin{figure}
\centering
\mbox{\epsfysize=12.cm\epsffile{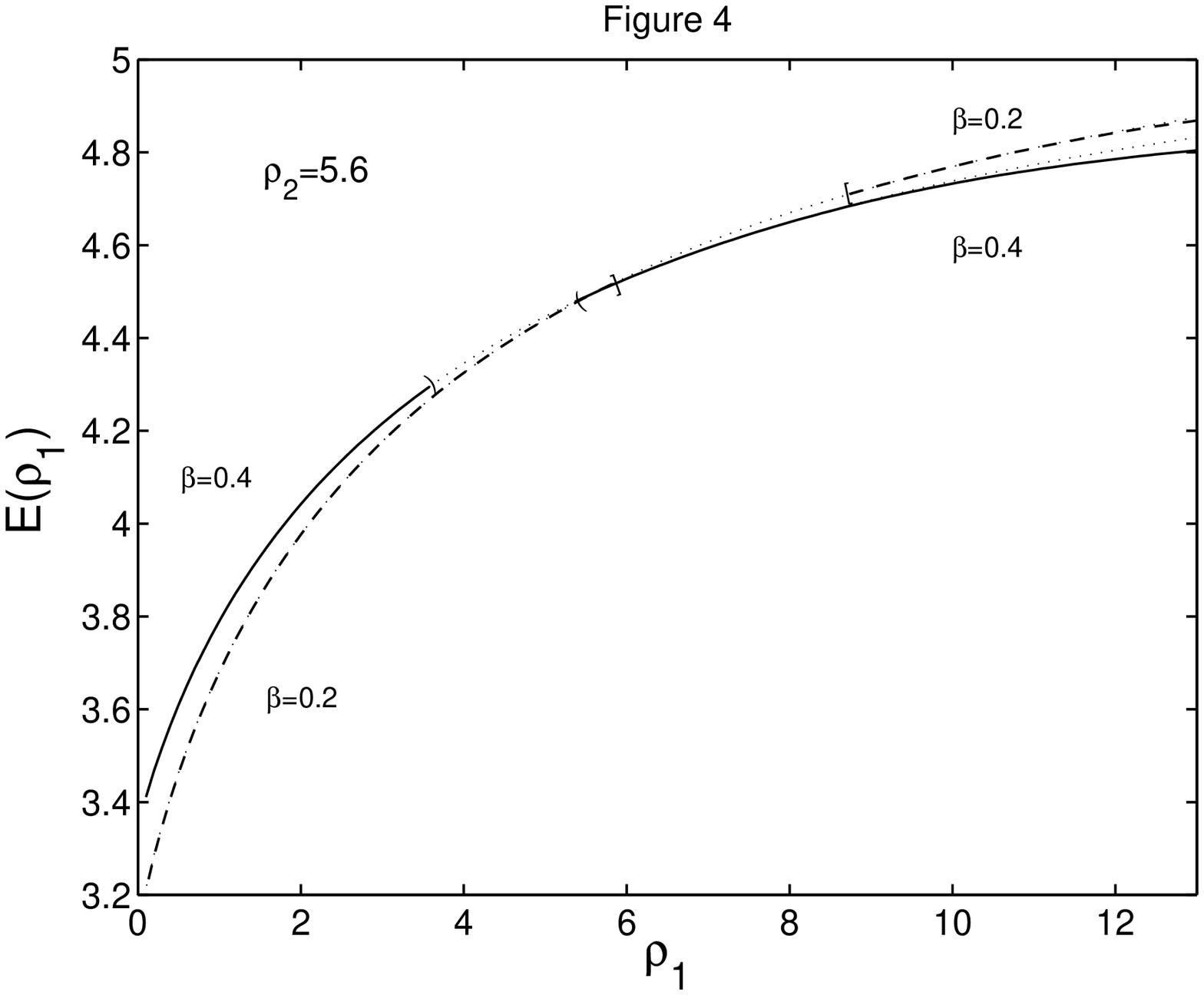}}
\end{figure}

\end{document}